\documentclass[10pt,A4paper,conference]{IEEEtran}
\usepackage{ifthen}
\newcounter{newctr}
\ifthenelse{\value{newctr}=1}{
\usepackage{latexsym}
\usepackage{graphicx}
\usepackage{dsfont}
\usepackage{amssymb}
\newtheorem{theorem}{Theorem}[section]
\newtheorem{lemma}{Lemma}[section]
\newtheorem{corollary}{Corollary}[section]
\newtheorem{discussion}{Discussion}[section]
\def\tw{.5\textwidth}
\def\QED{\mbox{\rule[0pt]{1ex}{1ex}}}
\def\Q{\hspace*{\fill}~\QED\par\endtrivlist\unskip}
\title{Stability of Scheduled Multi-access Communication over 
Quasi-static Flat Fading Channels with Random Coding and Independent
Decoding
}
\author{\authorblockN{KCV Kalyanarama Sesha Sayee, Utpal Mukherji}
\authorblockA{Dept. of Electrical Communication Engineering \\
Indian Institute of Science, Bangalore-560012, India \\
Email: sayee, utpal@ece.iisc.ernet.in} 
}}
{

\usepackage{psfig}
\usepackage{amsfonts}
\usepackage{dsfont}
\usepackage{fullpage}
\usepackage{color}
\input{srinadh.tex}
\textheight 9in
\textwidth 6.5in
\evensidemargin=0in
\oddsidemargin=0in
\topmargin= -.5in
\parindent 0.5in
\parskip 1.5ex
\newtheorem{lemma}{Lemma}[section]
\newtheorem{theorem}{Theorem}[section]

\newtheorem{corollary}{Corollary}[section]

\usepackage{graphicx}
\def\tw{\textwidth}
\def\QED{\mbox{\rule[0pt]{1ex}{1ex}}}
\def\Q{\hspace*{\fill}~\QED\par\endtrivlist\unskip}

\title{Stability of Scheduled Multi-access Communication over 
Quasi-static Flat Fading Channels with Random Coding and Independent
Decoding
}
\author{KCV Kalyanarama Sesha Sayee and Utpal Mukherji \\
Dept. of Electrical Communication Engineering \\
Indian Institute of Science, Bangalore-560012, India \\
Email: \texttt{[sayee, utpal]@ece.iisc.ernet.in} } 
}
\begin{document}

\maketitle

\begin{abstract}
The stability of scheduled multiaccess communication with random coding
and independent decoding of messages is investigated. The number of
messages that may be scheduled for simultaneous transmission is limited
to a given maximum value, and the channels from transmitters to receiver
are quasi-static, flat, and have independent fades. Requests for message
transmissions are assumed to arrive according to an i.i.d.  arrival
process.  Then, we show the following: (1) in the limit of large message
alphabet size, the stability region has an interference limited
information-theoretic capacity interpretation, (2) state-independent
scheduling policies achieve this asymptotic stability region, and (3) in
the asymptotic limit corresponding to immediate access, the stability
region for non-idling scheduling policies is shown to be identical
irrespective of received signal powers.
\end{abstract}

\section{Introduction}
The random-coding upper bound on mean message error probability for
maximum likelihood decoding in the presence of noise is exponentially
tight in the code length for rates sufficiently close to capacity.
Multi-access random-coded communication with independent decoding, of
messages that arrive in a Poisson process to an infinite transmitter
population, and that achieves any desired value for the upper bound by
determining message signal durations appropriately, has been considered
in~\cite{TelGal-JRN-JSAC} and~\cite{Tel-THESIS}.  In the present work
the number of messages that may be scheduled for simultaneous
transmission is limited to a given maximum value, and the channels from
transmitters to receiver are quasi-static, flat, and have independent
fades. Requests for message transmissions are assumed to arrive
according to an i.i.d.  arrival process.  Then, we show the following:
(1) in the limit of large message alphabet size, the stability region
has an interference limited information-theoretic capacity
interpretation, (2) state-independent scheduling policies achieve this
asymptotic stability region, and (3) in the asymptotic limit
corresponding to immediate access, the stability region for non-idling
scheduling policies is shown to be identical irrespective of received
signal powers.

First, we describe the model of~\cite{TelGal-JRN-JSAC}
and~\cite{Tel-THESIS} in brief.  Consider a multi-access message
communication system.  Requests for message transmissions over a flat
bandpass additive white Gaussian noise (AWGN) channel arrive according
to  Poisson process.  Messages are selected from a finite message
alphabet of size $M$. Each message has to be transmitted reliably with
reliability quantified by the tolerable message decoding error
probability, $P_{e}$.  Upon each message arrival the receiver assigns a
codebook of Gaussian signals with zero mean, equal power $P$ and uniform
power spectral density over a narrow frequency band of width $W$,
following random coding. The receiver uses the codebook of a transmitter
in independent maximum-likelihood decoding of the message of the
transmitter. Each transmitter transmits its signal, starting at its
message arrival time, for a random duration determined by the receiver.
This signalling duration of a message is chosen so that the random
coding bound on the expected probability of message decoding error
equals a pre-set value.  A noiseless feedback channel from the receiver
to the transmitters, and mechanism by which transmitters inform the
receiver of their intention to transmit are assumed.  In
~\cite{TelGal-JRN-JSAC} and~\cite{Tel-THESIS} this random-coded
multi-access system is then modelled as a processor-sharing queue in
which the transmitters are ``customers'' that are ``served'' by the decoder.
The processor-sharing model is then analyzed to determine the stability
condition and the mean delays experienced by the incoming messages, by
determining steady-state probabilities.

In the present work, we generalize this model. First, we limit the
maximum number of simultaneous message transmissions on the channel to a
finite positive integer $\mathsf{K}$. This assumption has the effect of
limiting the interference as seen by any message transmission, i.e.,
atmost $\mathsf{K}-1$ transmissions can interfere.  Second, we assume
independent quasi-static flat fading from the transmitters to the
receiver in the channels. With this assumption, there is an i.i.d.
multiplicative gain in the channel from each transmitter to the
receiver.  Thus, for a transmitted signal power $P_{m}$ for message $m$,
the received signal power is equal to $|h_{m}|^{2}P_{m}$, where the
multiplicative gain $h_{m}$, which is assumed to be known at the
receiver, is a random variable that has a finite number of finite
possible magnitudes. We assume that each message is assigned a message
class upon arrival, that is uniquely identified by the power level it
chooses from a finite set of power levels and by its service requirement

Further, we assume that the receiver schedules messages for transmission
in each Nyquist-sample-time-slot and for each system state based on
their message classes. We consider two classes of scheduling policies (
defined in sections~~\ref{section:state-dependent} and
~\ref{section:state-independent}): (1) non-idling policies , defined in
section~\ref{section:state-dependent} and denoted by
$\Omega_{\mathsf{K}}$, which are required to schedule as many as
possible up to $\mathsf{K}$ simultaneous message transmissions, and (2)
state-independent scheduling policies, defined in
section~\ref{section:state-independent} and denoted by
$\Omega^{\mathsf{K}}$, which are required to schedule not more than
certain numbers of transmissions of the various classes and a total of
\emph{not} more than $\mathsf{K}$ transmissions.

We derive inner and outer bounds to the stability region achievable by
scheduling policies in the classes $\Omega_{\mathsf{K}}$ and
$\Omega^{\mathsf{K}}$ policies. We show that in the limit $\mathsf{K}
\rightarrow \infty$, for scheduling policies in $\Omega_{\mathsf{K}}$,
the stability region characterization is independent of message SNR-s
and fading distribution. For scheduling policies in
$\Omega^{\mathsf{K}}$, we derive \emph{exact} characterization of the
nat arrival rate stability region in the limit of large message alphabet
sizes.  We also give interference limited information-theoretic capacity
interpretations to the stability regions for non-idling scheduling
policies in the equal powers case, and for state-independent scheduling
policies.

The organization of the paper is as follows. Section~\ref{section:model}
describes the model considered in the present work and also introduces
the notation.  Section~\ref{section:state-dependent} presents stability
analysis for scheduling policies in the class $\Omega_{\mathsf{K}}$.
Section~\ref{section:state-independent} presents stability analysis for
scheduling policies in the class $\Omega^{\mathsf{K}}$.
Section~\ref{section:outer bound} gives an outer bound to the stability
region achievable by any scheduling policy.

\section{Modeling and Analysis} \label{section:model}
We first extend the definition of service quantum
in~\cite{TelGal-JRN-JSAC} and~\cite{Tel-THESIS} so as to apply in our
extended model.  The equivalent baseband channel between the transmitter
for a message $m$ and the receiver can be sampled at the Nyquist rate
equal to its two-sided bandwidth $W$, to obtain a sequence of single-use
scalar channels. These channels $i$, when conditioned on knowledge of
the set $N(i)$ and coefficients $h_{n}$ of overlapping transmissions,
have independent inputs of variance $P_{m} =|h_{m}|^{2}P$ and
independent effective noises of variance $\sigma^{2}_{m,i} = N_{0}W+
\sum_{n \in N(i),\;\; n \neq m} |h_{n}|^{2}P$, where $P$ denotes the
power of a message signal before multiplication by its coefficient and
$N_{0}$ denotes the baseband noise two-sided power spectral density. The
random coding bound on the mean error probability in decoding message
$m$ from the outputs of the first $d$ scalar channels is as follows: for
any $0 < \rho \leq 1$,
\[
\overline{P_{e}} \leq \mathrm{exp}\left[\rho \ln M - \sum_{i=1}^{d}
E_{0}(\rho,P_{m}, \sigma^{2}_{m,i})\right],
\]
where $M$ denotes the size of the message alphabet and 
\[
E_{0}(\rho,P_{m},\sigma^{2}_{m,i}) = \rho
\ln\left(1+\frac{P_{m}}{(1+\rho)\sigma^{2}_{m,i}}\right).
\]
For a particular choice of $\rho$ and tolerable error probability
$P_{e}$, if $d$ is the smallest integer for which
\[
\sum_{i=1}^{d} E_{0}\left(\rho,P_{m},\sigma^{2}_{m,i}\right) \geq -\ln P_{e}+
\rho \ln M,
\]
then $\overline{P_{e}} \leq P_{e}$. Thus, $E_{0}(\rho, P_{m},
\sigma^{2}_{m,i})$ can be considered to be service quantum received by
message $m$ over scalar channel $i$, and $S = -\ln P_{e}+\rho \ln M$ can
be considered to be the service requirement of message $m$. The
corresponding signal duration for message $m$, measured in terms of
number of scalar channel uses or slots, is $d$.

Requests for message transmissions are assumed to arrive at slot
boundaries in batches.  Each message, upon arrival, is assigned a class
$(l, j)$ based on its service requirement and fading value, where $1
\leq l \leq L$ and $1 \leq j \leq J$, assuming that service requirement
takes $L$ different values and fading takes $J$ different values.  Let
$P_{e,l}$ and $M_{l}$ be the tolerable error probability and message
alphabet size for class-$(l, j)$ messages respectively.  A class-$(l,
j)$ message then has a service requirement equal to $S_{l} = -\ln
P_{e,l}+\rho \ln M_{l} < \infty$ and received power $P_{j} > 0$.
Throughout this paper, the suffix $l$ will identify service requirement
class and suffix $j$ the received power class.  Let the random variable
$A_{lj}$, with finite moments $\mathds{E}A_{lj}$ and
$\mathds{E}A_{lj}^{2}$, represent the number of messages of class-$(l,
j)$ that arrive in any slot, with the pmf $\Pr(A_{lj}=k) = p_{lj}(k),\;k
\geq 0$.  We assume that $\{A_{lj}\}$ are independent random variables.
Let $\mathds{E}A = (\mathds{E}A_{11}, \mathds{E}A_{12}, \ldots,
\mathds{E}A_{lj}, \ldots, \mathds{E}A_{LJ}) \in \mathbb{R}_{+}^{LJ}$.
Let $\lambda_{lj}$ denote the arrival rate of class-$(l, j)$ messages.
Since each slot is of duration $\frac{1}{W}$, we have $\lambda_{lj} = W
\mathds{E}A_{lj}$.  Let $\Gamma_{j} = \frac{P_{j}}{N_{0}W}$ be the
received-message-signal to noise ratio (\textsc{SNR}) of a message
received at power $P_{j}$.

We construct a discrete-time countable state space Markov-chain model
and then analyze for the stability ($c$-regularity~\cite{Mey-paper}) of
the model.  These stability results are derived by obtaining appropriate
drift conditions for suitably defined Lyapunov functions of the state of
the Markov chain. In particular, we prove that the Markov chain is
$c$-regular by applying Theorem 10.3 from~\cite{Mey-paper}, and then 
show finiteness of the stationary mean number of
messages in the system.

Thus, in the following sections we obtain inner and outer bounds to the
achievable stability region for scheduling policies in
$\Omega^{\mathsf{K}}$ and $\Omega_{\mathsf{K}}$. In some particular
cases we derive exact characterization of the stability region.

Let $s = (s_{11}, s_{12}, \ldots, s_{lj}, \ldots, s_{LJ}) \in
\mathbb{Z}_{+}^{LJ}$ be a vector of non-negative integers and define two
sets $\mathcal{S}_{\mathsf{K}}  = \left\{s: 0 \leq \sum_{l=1}^{L}
\sum_{j=1}^{J} s_{lj} \leq \mathsf{K} \right\} $, and
$\overline{\mathcal{S}}_{\mathsf{K}}  = \left\{s: \sum_{l=1}^{L}
\sum_{j=1}^{J} s_{lj} = \mathsf{K} \right\} \subset
\mathcal{S}_{\mathsf{K}}$ where $\overline{\mathcal{S}}_{\mathsf{K}}$
denotes the set of all schedules that schedule \emph{exactly}
$\mathsf{K}$ messages simultaneously for transmission and
$\mathcal{S}_{\mathsf{K}} $ denotes the set of all schedules that do not
schedule more than $\mathsf{K}$ transmissions.  We also define the
following subset of $\overline{\mathcal{S}}_{\mathsf{K}}$:
$\overline{\mathcal{S}}_{\mathsf{K}}(j) = \left\{s \in
\overline{\mathcal{S}}_{\mathsf{K}} : \sum_{l=1}^{L} s_{lj} > 0 \right\}$
denotes the set of schedules in $\overline{\mathcal{S}}_{\mathsf{K}}$
that schedule at least one class-$(.,j)$ message.  Let $\phi_{j}(s) > 0$
be the available service quantum per class-$(., j)$ message under the
schedule $s \in \overline{\mathcal{S}}_{\mathsf{K}}(j)$, and $\phi_{j}(s)
= 0$ otherwise.  Let $\underline{\phi}_{j} = \min_{s \in
\overline{\mathcal{S}}_{\mathsf{K}}(j)} \phi_{j}(s)$,
$\overline{\phi}_{j} = \max_{s \in
\overline{\mathcal{S}}_{\mathsf{K}}(j)} \phi_{j}(s)$, $\underline{\phi} =
\min_{s \in \overline{\mathcal{S}}_{\mathsf{K}}} \sum_{l=1}^{L}
\sum_{j=1}^{J} s_{lj} \phi_{j}(s)$.  The following notation will be used
in the rest of the paper: for any $x>0$ and $q>0$, $\lceil x \rceil_{q}
= \min (n \geq 1: x \leq nq)q$.

\section{Stability for Non-idling Scheduling Policies} 
\label{section:state-dependent}
Formally, a non-idling deterministic scheduling policy in the class
$\Omega_{\mathsf{K}}$ of policies, denoted by $\omega$, is defined by
the mapping $\left\{ \omega(\alpha): \mathcal{X} \rightarrow
\mathcal{S}_{\mathsf{K}} \right\}$.  We also define a non-idling random
scheduling policy $\omega$ in the class $\Omega_{\mathsf{K}}$ of
policies as a collection of random variables $\mathsf{U}_{\alpha}$,
$\alpha \in \mathcal{X}$. Here, $\mathsf{U}_{\alpha}$ takes values in
$\varphi(\alpha)$ with probability measure $p\left( \varphi(\alpha)
\right)$, where $\varphi(\alpha)$ denotes the set of schedules
implementable in state $\alpha$.  Throughout this paper we use the same
notation $\omega$ to denote both deterministic as well as random policy.
Formally, this class of scheduling policies is represented as $\left\{
\mathsf{U}_{\alpha}, p\left( \varphi(\alpha) \right), \alpha \in
\mathcal{X} \right\}$. Thus the set $\Omega_{\mathsf{K}}$ is composed of
both deterministic and random non-idling scheduling policies $\omega$.

Let $x$ denote the residual service requirement of a message of any
class. Define state as
\begin{eqnarray}
\label{eq:dependent state}
\alpha = \left( (x_{1}, l_{1}, j_{1}), (x_{2}, l_{2}, j_{2}),
\ldots, (x_{n(\alpha)}, l_{n(\alpha)}, j_{n(\alpha)}) \right) 
\end{eqnarray}
where $n(\alpha)$ is the number of messages in state $\alpha$, and
$(x_{k}, l_{k}, j_{k})$ gives the residual service requirement, service
requirement class, and received power class, respectively for the $k$th
message in state $\alpha$.  Let $\mathcal{X}$ be the countable set of
state vectors $\alpha$ defined in~(\ref{eq:dependent state}) over which
the discrete-time Markov chain $\{ X_{n}, n \geq 0 \}$ is defined.
Countability of the state space $\mathcal{X}$ follows from the fact that
the number of service quantums that a message of class-$(l, j)$ needs
for its service completion is bounded above by $\left\lceil
\frac{S_{l}}{\underline{\phi}_{j}} \right\rceil $. Let $V(\alpha)$ be a
Lyapunov function defined on $\mathcal{X}$ and  let $\mathcal{R}_{in}
\left(\Omega_{\mathsf{K}} \right)$ denote the set of message arrival
rate vectors $\mathds{E}A$ for which the Markov chain $\{ X_{n}, n \geq
0\} $ is positive recurrent and yields finite stationary mean for the
number of messages of each class for every $\omega \in
\Omega_{\mathsf{K}}$. Also, let $\mathcal{R}_{out}
\left(\Omega_{\mathsf{K}} \right)$ denote the set of $\mathds{E}A$ for
which the Markov chain is transient for every $\omega \in
\Omega_{\mathsf{K}}$. Then we prove the following two lemmas and two
theorems.

\begin{lemma} 
\label{lemma:unequal1}
Let $\mathsf{K} \geq 1$. For $\alpha \in \mathcal{X}$, let $c(\alpha)
= \sum_{k=1}^{n(\alpha)} \left\lceil \frac{x_{k}}{\underline{\phi}_{j_{k}}}
\right\rceil +1$ and 
\[
V(\alpha) = \frac{c^{2}(\alpha)}{2\left( \mathsf{K}
-\sum_{l=1}^{L} \sum_{j=1}^{J} \mathds{E}A_{lj} \left\lceil
\frac{S_{l}}{\underline{\phi}_{j}} \right\rceil \right)}. 
\]
Then for
$\omega \in \Omega_{\mathsf{K}}$ the Markov chain is $c$-regular if
$\sum_{l=1}^{L} \sum_{j=1}^{J} \mathds{E}A_{lj} \left\lceil
\frac{S_{l}}{\underline{\phi}_{j}} \right\rceil  < \mathsf{K}$.  
\end{lemma}

\begin{lemma}
\label{lemma:unequal2}
Let $\mathsf{K} \geq 2$. For $\alpha \in \mathcal{X}$,  let $c(\alpha)
= \sum_{k=1}^{n(\alpha)} \left( x_{k} + \overline{\phi}_{j_{k}} \right)
+1$ and \[ V(\alpha) = \frac{c^{2}(\alpha)}{2 \left( \underline{\phi} -
\sum_{l=1}^{L} \sum_{j=1}^{J}
\mathds{E}A_{lj}\left(S_{l}+\overline{\phi}_{j}\right) \right)}.
\]
Then for $\omega \in \Omega_{\mathsf{K}}$ the Markov chain is
$c$-regular if $
\sum_{l=1}^{L} \sum_{j=1}^{J}
\mathds{E}A_{lj}\left(S_{l}+\overline{\phi}_{j}\right) < \underline{\phi}$
\end{lemma}

\begin{theorem}
\label{th:unequal1}
Let $\mathsf{K}=1$. Then for $\omega \in \Omega_{\mathsf{K}}$ the Markov chain is 
\begin{itemize}
\item[(a)] positive recurrent and
yields finite stationary mean for the number of messages of each class if
$\sum_{l=1}^{L} \sum_{j=1}^{J} \mathds{E}A_{lj} \left\lceil
\frac{S_{l}}{\underline{\phi}_{j}}
\right\rceil < 1$ and, 

\item[(b)] transient if $
\sum_{l=1}^{L} \sum_{j=1}^{J} \mathds{E}A_{lj} \left\lceil
\frac{S_{l}}{\underline{\phi}_{j}}
\right\rceil > 1$.  
\end{itemize}
\end{theorem}

\begin{theorem}
\label{th:unequal2}
Let $\mathsf{K} \geq 2$. 
\begin{itemize}
\item[(a)] For $\omega \in \Omega_{\mathsf{K}}$, an arrival
rate vector $\mathds{E}A$ satisfying inequality ~(\ref{eq:pr3}) or
inequality~(\ref{eq:pr4}) below belongs to
$\mathcal{R}_{in} \left(\Omega_{\mathsf{K}} \right)$.

		\begin{eqnarray}
		\label{eq:pr3}
		\sum_{l=1}^{L} \sum_{j=1}^{J} \mathds{E}A_{lj} \left\lceil \frac{S_{l}}
		{\underline{\phi}_{j}} \right\rceil < 
		\mathsf{K} 
		\end{eqnarray}
		\begin{eqnarray}
		\label{eq:pr4}
		\sum_{l=1}^{L} \sum_{j=1}^{J} \mathds{E}A_{lj} \left(S_{l}
		+ \overline{\phi}_{j} \right)<
		\underline{\phi}
		\end{eqnarray}
		
\item[(b)]	For $\omega \in \Omega_{\mathsf{K}}$, for  every
		non-empty subset $B$ of $\{ 1, 2, \ldots, J \}$, an
		arrival rate vector $\mathds{E}A$ satisfying
		inequality~(\ref{eq:tran1}) below belongs to
		$\mathcal{R}_{out} \left(\Omega_{\mathsf{K}} \right)$.  
		\begin{eqnarray}
		\label{eq:tran1}
		\sum_{l=1}^{L} \sum_{j \in B} S_{l} \mathds{E}A_{lj} &\geq& 
		\max_{s \in \overline{\mathcal{S}}_{\mathsf{K}} }
		\sum_{l=1}^{L} \sum_{j \in B} s_{lj} \phi_{j}(s)
		\end{eqnarray} \Q
		
\end{itemize}
\end{theorem}

Part ($a$) of Theorem~\ref{th:unequal1} follows from
Lemma~\ref{lemma:unequal1}. To prove Part ($b$) we show that for the
Lyapunov function $V(\alpha) = 1 -\theta^{c(\alpha)}$, where $c(\alpha)$
is as defined in Lemma~\ref{lemma:unequal1}, there exists a value for
$\theta$, $0 < \theta < 1$, for which $V(\alpha)$ satisfies the
conditions for the theorem for transience~\cite{MeyTwe-BOOK}.
Inequalities~(\ref{eq:pr3}) and~(\ref{eq:pr4}) of
Theorem~\ref{th:unequal2} follow from Lemmas~\ref{lemma:unequal1}
and~\ref{lemma:unequal2} respectively. To prove
inequality~(\ref{eq:tran1}) of Theorem~\ref{th:unequal2} we proceed as
follows.  For a non-empty subset $B$ of $\{1, 2, \ldots, J\}$, define
$r_{B}(\alpha) = \sum_{k=1}^{n(\alpha)} x_{k} I_{\{j_{k} \in B\}}$,
where $I_{\{.\}}$ is the indicator function, and $V(\alpha) =
1-\theta^{r_{B}(\alpha)}$. We then show that for this Lyapunov function
there exists a value for $\theta$, $0 < \theta < 1$, for which
$V(\alpha)$ satisfies the conditions for the theorem for transience.

\begin{corollary}
\label{coro:unequal limit case}
In the limit $\mathsf{K} \rightarrow \infty$ the Markov chain $\{ X_{n},
n \geq 0 \}$ is
\begin{itemize}
\item[(a)]   positive recurrent and yields finite stationary mean for the number of
	messages if $\sum_{l=1}^{L} \sum_{j=1}^{J} \mathds{E}A_{lj} S_{l}<
        \frac{\rho}{1+\rho}$ , and
\item[(b)]   transient if $\sum_{l=1}^{L} \sum_{j=1}^{J} 
	\mathds{E}A_{lj} S_{l} > \frac{\rho}{1+\rho}$
\end{itemize} \Q
\end{corollary}

Corollary~\ref{coro:unequal limit case} follows from 
inequality~(\ref{eq:pr4}) of Theorem~\ref{th:unequal2} and 
inequality~(\ref{eq:tran1}) of the same theorem specialized for the set $B
= \{1, 2, \ldots, J\}$.  The corollary says that, in the limit
$\mathsf{K} \rightarrow \infty$, the stability result is
\emph{independent} of message SNR-s and their distribution.


The stability results for the \emph{continuous-time} models
in~\cite{TelGal-JRN-JSAC} and~\cite{UtpSan-CONF-ITW2002} coincide with
the corresponding results, stated in Corollary~\ref{coro:unequal limit
case}, for the discrete-time model in the limit of large number of
simultaneous transmissions.

In the remainder of this section we consider the special case $J=1$
(corresponding to \emph{no} fading) for which \emph{exact}
characterization of the stability region $\mathcal{R}_{in} \left(
\Omega_{\mathsf{K}} \right)$ can be given as follows.

\begin{corollary}
\label{coro:Equal Powers}
Let $J=1$ and $\mathsf{K} \geq 1$. For $\omega \in \Omega_{\mathsf{K}}$
the Markov chain $\{ X_{n}, n \geq 0 \}$ is
\noindent
\begin{itemize}
\item[(a)]   positive recurrent and yields finite stationary mean for the number of
	messages of each class if $\sum_{l=1}^{L} \mathds{E}A_{l1} \lceil S_{l}
	\rceil_{\underline{\phi}_{1}} < \mathsf{K} \underline{\phi}_{1},$ 
	and in the limit $\mathsf{K} \rightarrow \infty$, $\sum_{l=1}^{L} 
	\mathds{E}A_{l1}S_{l} < \frac{\rho}{1+\rho}$, and
\item[(b)]   transient if 
	$\sum_{l=1}^{L} \mathds{E}A_{l1} \lceil S_{l} \rceil_{\underline{\phi}_{1}} >
	\mathsf{K} \underline{\phi}_{1},$
	and in the limit $\mathsf{K}\rightarrow \infty$, $\sum_{l=1}^{L}
	\mathds{E}A_{l1}S_{l} > \frac{\rho}{1+\rho}$.
\end{itemize} \Q
\end{corollary}

Part ($a$) of Corollary~\ref{coro:Equal Powers} follows from
inequality~(\ref{eq:pr3}) of Theorem~\ref{th:unequal2}. To prove Part
($b$) we proceed as follows. Define $r(\alpha) = \sum_{k=1}^{n(\alpha)}
\lceil x_{k} \rceil_{\underline{\phi}_{1}}$ and $V(\alpha) =
1-\theta^{r(\alpha)}$. We then show that for this Lyapunov function
there exists a value for $\theta$, $0 < \theta < 1$, for which the
conditions for the theorem for transience are satisfied.

\begin{figure}
\centering
\includegraphics[width=\tw]{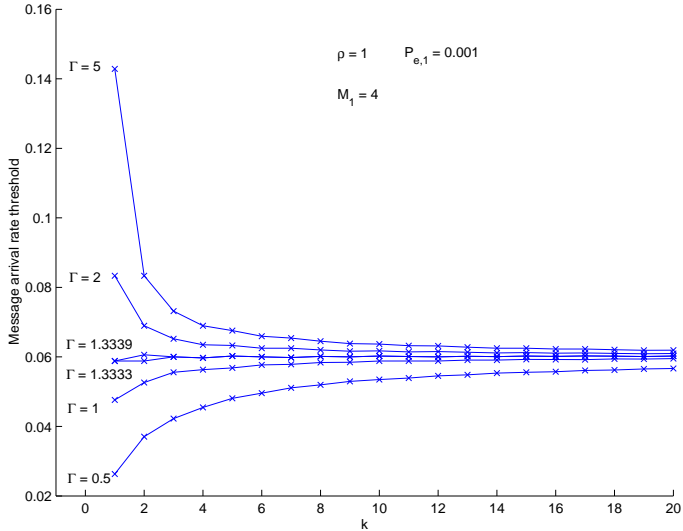}
\caption{Message arrival rate threshold versus maximum number of
simultaneous message transmissions, $\mathsf{K}$, in the case $J=L=1$.}
\label{fig:equal powers}
\end{figure}

Figure~\ref{fig:equal powers} shows plots of message arrival rate
stability limit versus $\mathsf{K}$, for the special case $J=L=1$ and
for different values of $\Gamma$ with parameters $\rho$, $M_{1}$ and
$P_{e, 1}$ fixed. From these plots we see that, for sufficiently small
transmit powers, as many simultaneous message transmissions as possible
should be scheduled, i.e., immediate access should be granted to
messages to increase the throughput of the system. For large transmit
power, scheduling many transmissions hurts the system throughput. This
behaviour can be explained as follows. For small transmit powers, the
effective noise seen by a transmission comes mainly from thermal noise,
rather than from interference caused by other ongoing message
transmissions.  Thus, interference from other signal transmissions has
insignificant effect on any given transmission, and scheduling as many
transmissions as possible is advantageous from the stability view point.
For large transmit powers, interference dominates the effective noise
seen by any message transmission.  Hence, limiting the number of
simultaneous transmissions is desirable.

Let
$f_{1}(M_{1}) = \frac{\mathsf{K} \underline{\phi}_{1} \ln M_{1}}{\lceil
S_{1} \rceil_{\underline{\phi}_{1}}}$ be the maximum stable nat
arrival rate in the special case $L=1$. Then the following corollary can
be proved.

\begin{corollary}[Capacity Interpretation]
\label{coro:capacity interpretation}
Let $L=J=1$. For a given $\Gamma$, $\mathsf{K}$, and $\rho$, the
threshold on nat arrival rate, $f_{1}(M_{1})$, increases with the
message alphabet size $M_{1}$ (in the sense that for any given positive
integer $M_{1}^{1}$ there exists a positive integer $M_{1}^{2} >
M_{1}^{1}$ such that $f_{1}(M_{1}^{2}) > f_{1}(M_{1}^{1})$), and
approaches the limit $\mathsf{K} W \ln \left( 1+
\frac{P}{(1+\rho)((\mathsf{K}-1)P+N_{0}W)} \right)$. As $\rho
\rightarrow  0$, this limit has its supremum that is equal to $\mathsf{K}$
times the information-theoretic capacity of an AWGN channel with SNR
$\frac{\Gamma}{(\mathsf{K}-1)\Gamma+1}$.   \Q 
\end{corollary}

\section{Stability for State-Independent Scheduling Policies}
\label{section:state-independent}
In this section we consider the class $\Omega^{\mathsf{K}}$ of
state-independent scheduling policies.  Formally, a policy in this class
is defined by a probability measure $ p(s)$, where $s \in
\mathcal{S}_{\mathsf{K}}$. To implement a scheduling policy $\omega$ in
$\Omega^{\mathsf{K}}$ we first classify incoming messages based on the
particular schedule $s$ to be assigned to them.

For each message arrival of class-$(l, j)$, a schedule $s \in
\left\{ s \in \mathcal{S}_{\mathsf{K}}: s_{lj} > 0 \right\}$ is chosen
randomly with some fixed probability measure and the message is further
classified by assigning the subclass-$(l, j, s)$ to it. With this
classification a message of subclass-$(l, j, s)$ will be scheduled to
transmit only when the schedule $s$ gets chosen.  We first fix a
scheduling policy $\omega =  p(s)$ and then, in each time slot, a
schedule $s$ is chosen from the set $\mathcal{S}_{\mathsf{K}}$,
\emph{independent} of the state $\alpha$, with probability $p(s)$. Let $x$ denote the residual
service requirement of a message of any class. Define state as 
\begin{eqnarray}
\label{eq:independent state}
\scriptstyle{\alpha
= \left( (x_{1}, l_{1}, j_{1}, z_{1}), (x_{2}, l_{2}, j_{2}, z_{2}),
\ldots, (x_{n(\alpha)}, l_{n(\alpha)}, j_{n(\alpha)}, z_{n(\alpha)})
\right)}
\end{eqnarray} 
where $n(\alpha)$ is the number of messages in state $\alpha$ and
$\left( x_{k}, l_{k}, j_{k}, z_{k} \right)$ gives the residual service
requirement, service requirement class, received power class, and the
assigned schedule, respectively, for the $k$th message in state $\alpha$.
When trying to implement a schedule $s$ the following two
possibilities can occur:

\begin{enumerate}
\item   there are enough messages of each subclass in state $\alpha$ to
	implement the chosen schedule , i.e., $n_{ljs}(\alpha) \geq
	s_{lj}$, where $n_{ljs}(\alpha)$ denotes the number of messages
	of subclass-$(l, j, s)$, in state $\alpha$.  Then we are able to
	completely implement the schedule.
\item   for at least one message subclass, the number of messages is less
        than that required by the schedule; then we transmit as many
	messages of such subclasses as are present in the system, and
	exactly as are required by the schedule in case of other
	subclasses. 
\end{enumerate}

Let $\phi_{j}(s)$ denote the available service quantum per class-$(., j,
s)$ message when schedule $s$ is \emph{completely} implemented.  Let
$\mathcal{X}$ be the countable set of state vectors $\alpha$ defined
in~(\ref{eq:independent state}) over which the discrete-time Markov
chain $\{ X_{n}, n \geq 0 \}$ is defined.  Countability of the state
space $\mathcal{X}$ follows from the fact that  the number of service
quantums that a message of subclass-$(l, j, s)$ needs for its service
completion is bounded above by $\lceil S_{l}/\phi_{j}(s) \rceil$.  Then
the following lemma and theorem are proved.

\begin{lemma}
Let $\mathsf{K} \geq 1$. For $\alpha \in \mathcal{X}$, and for each
subclass-$(l, j, s)$ let 
\[
c_{ljs}(\alpha) = \sum_{ \{k:\; (l_{k}, j_{k}, z_{k}) = (l, j, s) \}}
\lceil x_{k} \rceil_{\phi_{j}(s)}
\]
Define 
\[
c(\alpha) = 1 + \sum_{ljs} c_{ljs}(\alpha), \qquad \mbox{and}
\]
\[
V(\alpha) = \sum_{ljs}
\frac{c_{ljs}^{2}(\alpha)}{2\left( p(s)s_{lj}\phi_{j}(s) - \lceil
S_{l} \rceil_{\phi_{j}(s)} \mathds{E}A_{ljs} \right)}. 
\]
Then for $\omega \in \Omega^{\mathsf{K}}$ the Markov chain is
$c$-regular if, for each subclass-$(l, j, s)$, 
\[
\mathds{E}A_{ljs} < \frac{p(s)s_{lj}\phi_{j}(s)}{\lceil S_{l}
\rceil_{\phi_{j}(s)}}.
\] \Q
\end{lemma}

\begin{theorem}
Let $\mathsf{K} \geq 1$. For the scheduling policy $p(s) \in
\Omega^{\mathsf{K}}$, a sufficient condition that the Markov chain $\{
X_{n}, n \geq 0 \}$ is positive recurrent and yields finite stationary
mean for the number of messages of each subclass is: for $1 \leq l \leq
L$, $1 \leq j \leq J$, and $s \in \mathcal{S}_{\mathsf{K}}$ 
\[
\mathds{E}A_{ljs} < p(s) \frac{s_{lj} \phi_{j}(s)}{\lceil S_{l}
\rceil_{\phi_{j}(s)}},
\]
where $\mathds{E}A_{ljs}$ is the mean number of subclass-$(l, j, s)$ messages
that arrive in any slot. \Q
\end{theorem}
Define


\[
\psi_{lj} = \sum_{ \{ s \in \mathcal{S}_{\mathsf{K}}: s_{lj} >
0 \} } p(s)
\frac{s_{lj} \phi_{j}(s)}{\lceil S_{l} \rceil_{\phi_{j}(s)}}
\]
and the set 
\[
\mathcal{R}_{in}\left( \Omega^{\mathsf{K}} \right)  = \bigcup_{p(s)
\in \Omega^{\mathsf{K}}} \left\{ \beta \in \mathbb{R}_{+}^{LJ}:
\beta_{lj} \leq \psi_{lj} \right\}
\]

\begin{corollary}
\label{coro:in1}
For any given message arrival rate vector $\mathds{E}A \in
\mathcal{R}_{in}\left( \Omega^{\mathsf{K}} \right)$ there exists a
scheduling policy $p(s) \in \Omega^{\mathsf{K}}$ such that the Markov
chain is positive recurrent and yields finite stationary mean for the
number of messages of each class. \Q 
\end{corollary}

In the following corollary we give information-theoretic capacity interpretation of nat
arrival rate stability region.
\begin{corollary}[Capacity Interpretation]
For given $\Gamma$, $\mathsf{K}$, $\rho$, and $p(s) \in
\Omega^{\mathsf{K}}$, the threshold on nat
arrival rate of class-$(l, j)$ messages increases with the message
alphabet size $M_{l}$ and approaches the limit $\sum_{ \{s \in
\mathcal{S}_{\mathsf{K}}:s_{lj}>0 \} } p(s) \frac{s_{lj} \phi_{j}(s)}{\rho}$. As
$\rho \rightarrow 0$, this limit has its supremum that is equal to
\[
\sum_{s \in \mathcal{S}_{\mathsf{K}}:s_{lj}>0} p(s) s_{lj} W \ln \left( 1+
\frac{P_{j}}{\sum_{l=1}^{L} \sum_{j=1}^{J} s_{lj}P_{j} -P_{j} +N_{0}W}
\right).
\] \Q
\end{corollary}


\section{A General Outer Bound to the Stability Region}
\label{section:outer bound}
Consider message arrival processes $\{A_{lj}, 1 \leq l \leq L; 1 \leq j
\leq J \}$ and a stationary scheduling policy $\omega$ that does \emph{not} schedule more
than $\mathsf{K}$ simultaneous transmissions. 
Let $\pi(s)$ be a probability measure on $\mathcal{S}_{\mathsf{K}}$.
Define 
\[
\Psi_{lj} = \sum_{ \{ s \in
\mathcal{S}_{\mathsf{K}}: s_{lj} > 0 \} } \pi(s) \frac{s_{lj}
\phi_{j}(s)}{S_{l}}
\]
 and the set 

\[
\mathcal{R}_{out} =
\bigcup_{\pi(s)} \left\{ \beta \in \mathbb{R}_{+}^{LJ}:
\beta_{lj} \leq \Psi_{lj} \right\}
\]

\begin{theorem}
Let the Markov chain $\{ X_{n}, n \geq 0 \}$ be positive recurrent and
yield finite stationary mean for the number of messages in the system
for the message arrival processes $\{A_{lj}\}$ and the stationary
scheduling policy $\omega$.  Then $\mathds{E}A \in \mathcal{R}_{out}$.
\Q 
\end{theorem}

Define 
\[
\overline{\mathcal{R}}_{in}\left(
\Omega^{\mathsf{K}} \right)  = \bigcup_{p(s) \in \Omega^{\mathsf{K}}} 
\left\{ \overline{\beta} \in
\mathbb{R}_{+}^{LJ}: \overline{\beta}_{lj} \leq \overline{\psi}_{lj} \right\},
\]
where $\overline{\psi}_{lj} = \psi_{lj}\ln M_{l} $, and
\[
\overline{\mathcal{R}}_{out} = \bigcup_{\pi(s)}\left\{ \overline{\beta} \in
\mathbb{R}_{+}^{LJ}: \overline{\beta}_{lj} \leq \overline{\Psi}_{lj} \right\},
\]
where $\overline{\Psi}_{lj} = \Psi_{lj}\ln M_{l} $.  Then the following
corollary is proved.

\begin{corollary}
\label{coro:in2}
In the limit $\min_{1 \leq l \leq L} M_{l} \rightarrow \infty$ we have
$\overline{\mathcal{R}}_{in} \left( \Omega^{\mathsf{K}} \right) =
\overline{\mathcal{R}}_{out}$.  
\end{corollary} 



\section{Acknowledgment}
The authors gratefully acknowledge support from the DRDO-IISc Joint Program on
Mathematical Engineering.

\bibliographystyle{unsrt}
\bibliography{info}
\end{document}